\documentclass[useAMS, usenatbib]{mn2e}

\usepackage{graphicx}
\usepackage{amssymb}
\usepackage{color}

\newcommand{\nv}{}

\title[Non-equipartition regime]%
{Brightness temperature --- obtaining physical properties of non-equipartition plasma.}
\author[E.~E.~Nokhrina]
{E.~E.~Nokhrina$^{1}$\thanks{E-mail: nokhrina@phystech.edu (EEN)}
\\
$^{1}$Moscow Institute of Physics and Technology, Institutsky per.~9,
Dolgoprudny 141700, Russia\\
}

\begin{document}

\date{Accepted 2017 February 26; Received 2017 February 24; in original form 2016 December 11}

\pagerange{\pageref{firstpage}--\pageref{lastpage}} \pubyear{2017}

\maketitle

\label{firstpage}

\begin{abstract}
The limit on the intrinsic brightness temperature, attributed to `Compton catastrophe', has been established being $10^{12}$~K.
Somewhat lower limit of the order of $10^{11.5}$~K is implied if we assume that the radiating plasma is in equipartition 
with the magnetic field --- the idea that explained why the observed cores of active galactic nuclei sustained the limit lower than the `Compton catastrophe'.
Recent observations with unprecedented high resolution by the RADIOASTRON revealed systematic exceed in the observed brightness temperature.
We propose means of estimating the degree of non-equipartition regime in AGN cores. Coupled with the core-shift measurements the method allows us
to estimate independently the magnetic field strength and the particles number density  at the core. We show that the ratio of magnetic energy
to radiating plasma energy is of the order of $10^{-5}$, which means the flow in the core is dominated by the particle energy. 
We show that the magnetic field obtained 
by the brightness temperature measurements may be underestimated. We propose for the relativistic jets with small viewing angles the
non-uniform MHD model, and obtain the expression for the magnetic field amplitude about two orders higher than that for the uniform model. These
magnetic field amplitudes are consistent with the limiting magnetic field suggested by the `magnetically arrested disk' model.   
\end{abstract}


\section{Introduction}
\label{s:intro}

The previous observations of active galactic nuclei (AGN) in radio band all have limited
the cores brightness temperature by $10^{12}$~K. This phenomenon has been explained in \citet{Kell-PT-69}
as being an outcome of so called ``inverse Compton catastrophe''.

It can be illustrated by such an argument \citep{Saas-Fee-94}. Suppose we have an electron \nv{moving in
magnetic field $B$ with the velocity $v$, $\beta=v/c$, $c$ is a speed of light, and} with \nv{corresponding} Lorentz factor $\gamma$. It radiates synchrotron radiation with the power (see, e.g., \citet{R-L-79})
\begin{equation}
P_{\rm S}=\frac{4}{3}\sigma_{\rm T}c\beta^2\gamma^2U_{\rm B}.
\end{equation}   
Here $\sigma_{\rm T}=8\pi/3r_0^2$ is Thomson cross-section, $r_0=e^2/mc^2$ is an electron classical radius, \nv{$e$ and $m$ are electron charge and mass, $c$ is a speed of light}, and
$U_{\rm B}=B^2/8\pi$ is the magnetic energy density.  
The same electron loses its energy undergoing the inverse Compton scattering of photons, the power being
\begin{equation}
P_{\rm C}=\frac{4}{3}\sigma_{\rm T}c\beta^2\gamma^2U_{\rm ph},
\end{equation} 
where the photon energy density
\begin{equation}
U_{\rm ph}=\int\epsilon dn(\epsilon),
\end{equation}
with photon energy distribution $n(\epsilon)$.
The photon energy density \nv{$U_{\rm ph}$} comprises of synchrotron photons $U_{\rm ph0}$, once Comptonized photons $U_{\rm ph1}$, 
and so forth. The full power of Compton losses is described by \citet{Saas-Fee-94}
\begin{equation}
U_{\rm ph}=U_{\rm ph0}\left[1+\frac{U_{\rm ph0}}{U_{\rm B}}+\left(\frac{U_{\rm ph0}}{U_{\rm B}}\right)^2+...\right]=
\frac{U_{\rm ph0}}{1-U_{\rm ph0}/{U_{\rm B}}}.
\end{equation}
If $U_{\rm ph0}={U_{\rm B}}$, the power $P_{\rm C}$ diverges, which is referred to by the ``inverse Compton catastrophe''.

The result, obtained by \citet{Kell-PT-69} is the limiting brightness temperature $10^{12}$~K, beyond which the
``inverse Compton catastrophe'' takes place. The question was how does the source ``know'' this limit and
sustains its brightness temperature below the limit. The answer has been proposed by \citet{Readhead-94}. There is 
another limit on the brightness temperature --- the so called equipartition temperature. If radiating plasma and magnetic field
are in energy equipartition in the source, the corresponding temperature $T_{\rm eq}$ is just below the limiting by the
inverse Compton catastrophe one. 

However, recent observations of AGN radio cores with the high-resolution {\it RADIOASTRON}
program \citet{Kovalev-2016} questioned the existence of such a limit, since there are observations that systematically show
the brightness temperatures greater than not only $T_{\rm eq}$, but also $10^{12}$~K.

In this work we do not approach a question what is the physical process underlying such extreme brightness temperatures. 
We address the question of obtaining the non-equipartition physical parameters of the radiating domain such as the magnetic field $B$,
particle number density $n$ and the measure of non-equipartition $\Sigma$. This is an important issue. 
Indeed, the analytical and numerical modeling \citep{BN-06, Kom-07, Lyub-09, Tch-08, Tch-09} support the idea,
that relativistic jets from AGN must be in equipartition regime. In particular, the magnetization parameter 
\begin{equation}
\sigma=\frac{B^2}{4\pi nmc^2\Gamma^2}, \label{sigma1}
\end{equation}
which
is the ratio of Poynting vector flux to plasma kinetic energy flux, must be unity. Here $\Gamma$ is the bulk Lorentz 
factor of a jet, and $n$ is a proper particle number density. However, it is believed that only the small portion of particles radiate. 
Indeed, the bulk plasma velocity,
dictated by the magnetohydrodynamics, is exactly the drift velocity in crossed electric and magnetic fields.
So, a cold plasma, moving with the drift velocity, does not radiate. To produce the radiation, cold plasma must be disturbed and accelerated, the power law
\begin{equation}
dn=k_{\rm e}\gamma^{-p}d\gamma,\quad \gamma\in[\gamma_{\rm min};\,\gamma_{\rm max}],\label{dn1}
\end{equation}
being used to describe the energy distribution of the radiating plasma. \nv{Here $k_{\rm e}$ is an amplitude of the electron
energy distribution.} 
To model the radiation of particles in the magnetic field, we employ synchrotron emission and self-absorption.
Indeed,  spectral energy distributions (SED) performed for blazars (see e.g. \citet{Abdo-11}) demonstrate that
at the low energy the Compton part of the radiation does not play an important role. 

As to acceleration process itself, there are two main processes which may account for it. One is the particle acceleration
on shocks. This process can hardly account for the observed radiation, since it has been shown (see, e.g., \citet{Saas-Fee-94}), that
the acceleration is not efficient for the magnetized shocks. As the possibility for a particle once accelerated to return to the shock front
is suppressed, the Fermi acceleration mechanism does not work. The second process is the reconnection of magnetic field. It accelerates about 
one per cent of particles very effectively \citep{Sironi-13}, producing a power-law spectrum, with maximum particles energy
growing with the time of numerical simulation. 

\section{Jet parameters in Blandford--K\"onigl model}
\label{s:BK}

The following model is used to explain the properties of compact bright features observed in radio band (see \citet{Gould-79, LB-85, Lobanov-98, Zdz-15}): 
the radiation domain is either a uniform ``plasmoid'' or a uniform excited part of a continuous jet. The position of this radiating spherical \citep{Gould-79} domain along the jet $r$ defines the amplitudes of the particle number density and of the magnetic field (with uniform distribution across the radiating domain) according to Blandford--K\"onigl model \citep{BK-79} as
\begin{equation}
B(r)=B_0\left(\frac{r_0}{r}\right),\quad k_{\rm e}(r)=k_{\rm e,0}\left(\frac{r_0}{r}\right)^2.\label{BK_model}
\end{equation}
\nv{Here $B_0$ and $k_{\rm e,0}$ are magnetic field and particle number density amplitude at a distance $r_0$.}

The model has been used to obtain such physical parameters of jets as a magnetic field, a particle number density \citep{Lobanov-98, Hirotani-05, OSG-09},
and multiplicity parameter and Michel's \citep{Michel-69} magnetization parameter \citep{NBKZ-15} using a core-shift effect --- 
the observed shift of position of radio cores at different frequencies. All these result are based on the equipartition assumption in different formulation ---
either the energy densities of magnetic field and radiating particles are equal, or the fluxes of Poynting vector and total particle kinetic energy are equal
with the number of radiating particles consisting about 1 per cent of the total particle number density.   

\subsection{Magnetic field}
\label{ss:mf}

The observed flux, or observed brightness temperature, can be used to estimate the magnetic field in the radiating domain \citep{Zdz-15}.
The observed spectral flux $S_{\nu}$ of a core at the frequency $\nu$ can be expressed on the one hand through the brightness temperature $T_{\rm b}$ as
\begin{equation}
S_{\nu}=\frac{2\pi\nu^2\theta^2}{c^2}k_{\rm B}T_{\rm b},\label{S1}
\end{equation}
where $\theta$ is the angular size of a radiating domain. On the other hand the flux for the optically thick uniform source \nv{of radius $R$
at the distance $d$} can be written using the 
spectral photon emission rate $\rho_{\nu}$ and effective absorption coefficient $\ae_{\nu}$ as \citep{Gould-79}:
\begin{equation}
S_{\nu}=\pi\hbar\nu\frac{\rho_{\nu}}{\ae_{\nu}}\frac{R^2}{d^2}u(2R\ae_{\nu}),\label{S2}
\end{equation}
\nv{and the function of the optical depth $u(2R\ae_{\nu})$ is defined in \citet{Gould-79}}. The emission and absorption coefficients for the synchrotron-self-Compton model
can be written in a jet frame (primed), i.e. in a frame where the electric field vanishes, as \citep{GS-64, BG-70, Gould-79}:
\begin{equation}
\rho'_{\nu'}=4\pi\left(\frac{3}{2}\right)^{(p-1)/2}a(p)\alpha k'_{\rm e}\left(\frac{\nu'_{\rm B'}}{\nu'}\right)^{(p+1)/2},\label{rho1}
\end{equation}
\begin{equation}
\ae'_{\nu'}=c(p)r_0^2 k'_{\rm e}\left(\frac{\nu_0}{\nu'}\right)\left(\frac{\nu'_{\rm B'}}{\nu'}\right)^{(p+2)/2}.\label{ae1}
\end{equation}
\nv{Here $\nu'_{\rm B'}=eB'/mc$ is a gyrofrequency in the fluid frame, $\hbar$ is the Planck constant, $\alpha=e^2/\hbar c$ is the fine structure constant, and the functions $a(p)$ and $c(p)$ of the electron
distribution spectral index $p$ are defined in \citep{Gould-79}.} 

Equations (\ref{S1}) and (\ref{S2}) are written in an observer frame. However, the spectral flux is calculated in a jet (primed) frame using 
(\ref{rho1}) and (\ref{ae1}), where it is expressed as a function of a frequency $\nu'$, magnetic field $B'$ and particle number density amplitude $k'_{\rm e}$
in the jet frame. In order to rewrite a flux and a brightness temperature into observer frame we use the Lorentz invariant \citep{R-L-79}
$S_{\nu}/\nu^3$. To express a magnetic field and a particle number density in the nucleus frame, and a frequency in an observer frame,
we use the following relations. 
\nv{A particle number density amplitude $k'_{\rm e}$ in a fluid frame correlates with its value $k_{\rm e}$ in the nucleus frame as}
\begin{equation}
k'_{\rm e}=k_{\rm e}/\Gamma,
\end{equation}
an observed frequency transforms from the fluid frame into observer frame as
\begin{equation}
\nu'=\nu_{\rm obs}\frac{1+z}{\delta},\label{nu_nupr}
\end{equation}
and a brightness temperature $T_{\rm b,\,obs}=T_{\rm b}\delta/(1+z)$. Here
$z$ is a cosmological red-shift of a source and a Doppler factor of a flow $\delta=\left[\Gamma(1-\beta\cos\varphi)\right]^{-1}$.  
The viewing angle of a jet is $\varphi$.
\nv{We assume that the toroidal component of a magnetic field dominates the jet radiating region outside the light cylinder with its position defined by $R_{\rm L}=c/\Omega_{\rm F}$. Indeed, the MHD analytical \citep{Beskin-97, Narayan-07, Tch-08, Lyub-09, NBKZ-15} and numerical \citep{BTch-16} models provide that $B_{\varphi}\approx B_{\rm P}r/R_{\rm L}$. Thus, the magnetic fields transforms from the fluid frame into the nucleus frame as $B\approx B'\Gamma$.}

Equating the right-hand sides of equations (\ref{S1}) and (\ref{S2}), we obtain for the magnetic field 
\begin{equation}
B=k_0(p)\frac{m^3c^5}{e}\frac{\Gamma\delta}{1+z}\nu_{\rm obs}\left(k_{\rm B}T_{\rm b,\,obs}\right)^{-2},\label{Beq1}
\end{equation}
where the numerical factor $k_0$ depends on the electrons spectral index and is equal to
\begin{equation}
k_0(p)=3.6\cdot 10^{-1},\quad p=2.
\end{equation}
\nv{Particular spectral index $p$ may be found by fitting the jet spectrum for a particular source. 
From the theoretical point of view, it depends on the non-thermal mechanism of particle acceleration in AGN, which is under debate. 
The first-order Fermi mechanism 
working at shocks provides $p=2$ \citep{BO-78}. However, the numerical simulations demonstrate that this mechanism works effectively for low magnetization
flows \citep{Sironi-11}, with $p\approx 2.5$. On the other hand, magnetic reconnection provides means for particle acceleration
and formation of power-law spectrum with the spectral index $p$ depending on the flow magnetization \citep{Sironi-14} and ranging from $1.5$ to $4$.
Having in mind the uncertainty of the spectral index, we choose the value $p=2$ as a fiducial parameter characterizing the non-thermal spectrum of radiating particles.}
Substituting $p=2$ we obtain
\begin{equation}
\left(\frac{B}{\rm G}\right)=7.4\cdot 10^{-4}\frac{\Gamma\delta}{1+z}\left(\frac{\nu_{\rm obs}}{\rm GHz}\right)\left(\frac{T_{\rm b,\,obs}}{10^{12}{\rm K}}\right)^{-2}.\label{Eq0}
\end{equation} 
The radio core is observed at the peak spectral flux, with $\nu_{\rm obs}=\nu_{\rm peak}$ for 
the magnetic field $B$ and radiating particle number density $n$ at the surface of the optical depth
equal to unity. For each frequency the position of this surface is different (core-shift effect, see e.g. \citep{Lobanov-98}), so the 
defined by equation~(\ref{Eq0}) magnetic field is for the particular position $r_{\rm core}$ of the observed core.
The equation~(\ref{Eq0}) does not give us a full information about the magnetic field amplitude, since we do not know
the position of the core at the observed frequency. In order
to obtain the core position, we need the measurements of the core-shift effect \citep{Lobanov-98, Sok-11, Push-12} as well. 

\subsection{Measure of equipartition}
\label{s:mag}

The core-shift effect is a change in the observed position of a core at different frequencies \citep{Lobanov-98, OSG-09}.
It is connected with the self-absorption of the synchrotron sources (see e.g. \citet{Gould-79}): due to absorption we observe the surface
of the optical thickness equal to unity. Both synchrotron emission rate and the absorption depend on the emitting particle number
density and magnetic field magnitudes and distributions. The `stardart' core-shift formula by \citet{Lobanov-98} has been obtained under
certain assumptions: the Blandford-Konigl field and particle number density dependence on $r$ (\ref{BK_model}) and the equipartition
between the radiating plasma and magnetic field. The last assumption has been essential for the results, since measurements of the core-shift
allows only to estimate the dependence of magnetic field magnitude on particle number density.
The same equipartition assumption has been used to establish the `equipatition brightness temperature' by \citet{Readhead-94}.   
However, recent observations of the brightness temperature at high resolution provided by RADIOASTRON exceed this `equipartition' limit,
so, as has been indicated by \citet{Gomez-16}, there is, probably, no equipartition in a jet.  
However, the measurements of both brightness temperature and core-shift provides us with the instrument to estimate
the magnetic field and particle number density independently \citep{Zdz-15}, and thus obtain the measure of `non-equipartition'.   

Let us introduce the radiation magnetization 
$\Sigma$ --- the ratio of Poynting flux to
radiating particle energy flux. Each particle internal energy is given by
$mc^2\gamma'$, where $\gamma'$ is the Lorentz factor of radiating particles with respect to plasma bulk motion.
The radiating particle number density $n'_{\rm rad}$ is given in jet bulk motion proper frame.
The amplitude $k'_{\rm e}$ is defined by the radiating particles number density $n'_{\rm rad}$ depending on the 
magnitude of exponent $p$ as $n'_{\rm rad}=k'_{\rm e}f(p)$ with
\begin{equation}
f(p)=\left\{
\begin{array}{rl}
\displaystyle\frac{1}{1-p}\left(\gamma_{\rm max}^{1-p}-\gamma_{\rm min}^{1-p}\right),&\quad p\ne 1,\\ \ \\
\displaystyle \ln\frac{\gamma_{\rm max}}{\gamma_{\rm min}},&\quad p=1.
\end{array}
\right.\label{fp}
\end{equation} 
We assume $p\in(1;\,2]$, and the Lorentz factor of plasma in nucleus frame is defined by $\gamma=\gamma'\Gamma$.
In this case, the magnetization of radiating particles is
\begin{equation}
\Sigma=\frac{\Gamma(2-p)B^2f(p)}{4\pi mc^2n_{\rm rad}\left(\gamma_{\rm max}^{2-p}-\gamma_{\rm min}^{2-p}\right)}
\end{equation} 
for $p\ne 2$, and
\begin{equation}
\Sigma=\frac{\Gamma B^2f(p)}{4\pi mc^2 n_{\rm rad}\ln\frac{\gamma_{\rm max}}{\gamma_{\rm min}}}
\end{equation} 
for $p=2$.
Here $n_{\rm rad}$ is given in nucleus frame. 
For the equipartition between magnetic field and radiating particles $\Sigma=1$. We will be interested in
estimates for $\Sigma$ from the observations. This will allow us to connect the radiation magnetization $\Sigma$
with the physical properties of the radiating plasma. We will use the non-dimensional function $F_{\Sigma}(p)$ such as
\begin{equation}
\Sigma=\frac{\Gamma B^2}{mc^2 n_{\rm rad}}F_{\Sigma}(p).\label{Seq1}
\end{equation}

The expression connecting \nv{the jet physical parameters $B$ and $r_{\rm rad}$ with the position of radiating region $r$ and the 
observed frequency $\nu_{\rm obs}$} has been obtained by \citet{Lobanov-98, Hirotani-05, NBKZ-15}:
\begin{equation}
B^{2+p}n_{\rm rad}^2=\nu_{\rm obs}^{4+p}F_1^{-1}F_2^{-1}r^{-2},\label{cseq1}
\end{equation}
where coefficients 
\begin{equation}
F_1=\frac{c^2(p)(p-1)^2}{5(4+p)}\frac{e^4}{m^2c^2}\left(\frac{e}{2\pi mc}\right)^{2+p},
\end{equation}
and
\begin{equation}
F_2=\left(\frac{\delta}{\Gamma(1+z)}\right)^{4+p}\left(\frac{2\chi}{\delta\sin\varphi}\right)^2.
\end{equation}
Here $\chi$ is a jet half-opening angle for the conical jet.
Using (\ref{Seq1}), we rewrite $n_{\rm rad}$ as a function of $\Sigma$ and $B$, and substituting (\ref{Beq1}) into (\ref{cseq1})
we obtain 
the expression for the flow magnetization in a radiating domain as a function of its position $r$ from the central
source, the observed brightness temperature, an observed frequency, and geometrical and velocity factors:
\nv{  
\begin{equation}
\begin{array}{l}
\displaystyle\Sigma=4.1\cdot 10^{3}\left(1.7\cdot 10^2\right)^{-p}C_{\Sigma}(p)\frac{2\chi\Gamma^2}{\delta\sin\varphi}\left(\frac{\delta}{1+z}\right)^{p+5}\times \\ \ \\
\displaystyle\times\left(\frac{r}{\rm pc}\right)\left(\frac{\nu_{\rm obs}}{\rm GHz}\right)\left(\frac{T_{\rm b,\,obs}}{10^{12}{\rm K}}\right)^{-(p+6)}.
\end{array}\label{Seq2_0}
\end{equation}
Here 
\begin{equation}
\begin{array}{l}
\displaystyle C_{\Sigma}(p)=\frac{F_{\Sigma}(p)}{f(p)}\frac{c(p)}{\sqrt{5(4+p)}}(2\pi)^2\times \\ \ \\
\displaystyle\times\left[2.8(1.5)^{(p-1)/2}\frac{a(p)}{c(p)}\right]^{p+6}.
\end{array}
\end{equation}
For $p=2$ we obtain}
\begin{equation}
\begin{array}{l}
\displaystyle\Sigma=1.58\cdot 10^{-5}\frac{2\chi\Gamma^2\delta^6}{\sin\varphi(1+z)^7}\frac{F_{\Sigma}(2)}{f(2)}\times \\ \ \\
\displaystyle\times\left(\frac{r}{\rm pc}\right)\left(\frac{\nu_{\rm obs}}{\rm GHz}\right)\left(\frac{T_{\rm b,\,obs}}{10^{12}{\rm K}}\right)^{-8}.
\end{array}\label{Seq2}
\end{equation}

The expression above has been obtained assuming (i) the radiating \nv{region} has uniform distribution of $n_{\rm rad}$ and $B$; (ii) the radiating domain is optically thick; 
(iii) the jet is conical with half-opening
angle $\chi$, so that the jet geometrical thickness along the line of sight depends on $r,\;\chi$ and $\varphi$ (see \citet{Hirotani-05} for details); 
(iv) we observe the surface of the optical depth approximately equal to unity at the observed frequency $\nu_{\rm obs}$. This allows us to estimate the
order of $\Sigma$, assuming that $r$ is of the order of a parsec.

However, if we additionally adopt the Blandford--K\"onigl scalings for the magnetic field $B$ and particle number density $n_{\rm rad}$ (\ref{BK_model}), 
we will be able to correlate the position of a radiating domain $r$ with the observed frequency $\nu_{\rm obs}$. Indeed, 
substituting (\ref{BK_model}) into (\ref{cseq1}) one obtains the classical expression $\nu_{\rm obs}r$ proportional to the
physical parameters of a jet. The last conclusion is supported by multifrequency observations by \citet{Sok-11}. 
Thus, if we have, in addition to the measurement of the brightness temperature, the core-shift measurement, we can use it to obtain the radiating domain 
position.
As
\begin{equation}
r\sin\varphi=\theta_{\rm d}\frac{D_{\rm L}}{(1+z)^2},
\end{equation}
where $D_{\rm L}$ is a luminosity distance, we introduce 
\begin{equation}
\Delta \theta_{\rm d}=\Phi\left(\frac{1}{\nu_1}-\frac{1}{\nu_2}\right).\label{Eq6}
\end{equation}
With $\Delta\theta_{\rm d}$ being measured for the two frequencies $\nu_1$ and $\nu_2$, we can calculate $\Phi$ in ${\rm mas\;GHz}$  
and find the observed position of the core at given frequency as
\begin{equation}
r_{\rm core}=\frac{\Phi D_{\rm L}}{\nu_{\rm obs}\sin\varphi\,(1+z)^2}.
\end{equation}

Knowing the core-shift we can estimate the radial distance of the observed radiating domain of a jet:
\begin{equation}
\frac{r_{\rm obs}}{\rm pc}=\frac{4.8}{\sin\varphi(1+z)^2}\left(\frac{\nu_{\rm obs}}{\rm GHz}\right)^{-1}\left(\frac{\Phi}{\rm mas\cdot GHz}\right)\left(\frac{D_{\rm L}}{\rm Gpc}\right),\label{R4}
\end{equation}
and, consequently, the magnetization in the observed core as
\nv{
\begin{equation}
\begin{array}{l}
\displaystyle\Sigma=2.1\cdot 10^{4}\left(1.7\cdot 10^2\right)^{-p}C_{\Sigma}(p)\frac{2\chi\Gamma^2\delta^{p+4}}{\sin^2\varphi(1+z)^{p+7}}\times \\ \ \\
\displaystyle\times\left(\frac{D_{\rm L}}{\rm Gpc}\right)\left(\frac{\Phi}{\rm mas\cdot GHz}\right)\left(\frac{T_{\rm b,\,obs}}{10^{12}{\rm K}}\right)^{-(p+6)}.
\end{array}\label{Seq3_0}
\end{equation}
For $p=2$ the expression is}
\begin{equation}
\begin{array}{l}
\displaystyle\Sigma=7.7\cdot 10^{-5}\frac{2\chi\Gamma^2\delta^6}{\sin^2\varphi(1+z)^9}\frac{F_{\Sigma}(2)}{f(2)}\times \\ \ \\
\displaystyle\times\left(\frac{D_{\rm L}}{\rm Gpc}\right)\left(\frac{\Phi}{\rm mas\cdot GHz}\right)\left(\frac{T_{\rm b,\,obs}}{10^{12}{\rm K}}\right)^{-8}.
\end{array}\label{Seq3}
\end{equation}

\subsection{Radiating particles number density}
\label{ss:rpp}

In order to obtain the radiating particle number density in a radiating domain, we substitute (\ref{Beq1}) into (\ref{cseq1}):
\nv{
\begin{equation}
\begin{array}{l}
\displaystyle \left(\frac{n_{\rm rad}}{{\rm cm}^{-3}}\right)=1.1\cdot 10^{-3}(1.7\cdot 10^2)^p C_{\rm n}(p)\times \\ \ \\
\displaystyle\times\frac{\Gamma\sin\varphi(1+z)^{p+3}}{2\chi\delta^{p+2}}\left(\frac{r}{\rm pc}\right)^{-1}\left(\frac{\nu_{\rm obs}}{\rm GHz}\right)\left(\frac{T_{\rm b,\,obs}}{10^{12}{\rm K}}\right)^{p+2}.
\end{array}\label{neq2_0}
\end{equation}
Here
\begin{equation}
\displaystyle C_{\rm n}(p)=f(p)\frac{\sqrt{5(p+4)}}{c(p)}\left[2.8(1.5)^{(p-1)/2}\frac{a(p)}{c(p)}\right]^{-(p+2)}.
\end{equation}
For $p=2$ we obtain the estimate for radiating particles number density at the region with the position $r$:}
\begin{equation}
\begin{array}{l}
\displaystyle\left(\frac{n_{\rm rad}}{{\rm cm}^{-3}}\right)=4\cdot 10^{4}\frac{\Gamma\sin\varphi(1+z)^5}{2\chi\delta^4}f(2)\times \\ \ \\
\displaystyle\times\left(\frac{r}{\rm pc}\right)^{-1}\left(\frac{\nu_{\rm obs}}{\rm GHz}\right)\left(\frac{T_{\rm b,\,obs}}{10^{12}{\rm K}}\right)^{4}.
\end{array}\label{neq2}
\end{equation}

Using (\ref{R4}), one can obtain the expression for $n_{\rm rad}$ as a function of the observables
\nv{
\begin{equation}
\begin{array}{l}
\displaystyle\left(\frac{n_{\rm rad}}{{\rm cm}^{-3}}\right)=2.3\cdot 10^{-4}(1.7\cdot 10^2)^p C_{\rm n}(p)\times \\ \ \\
\displaystyle\times\frac{\Gamma\sin^2\varphi(1+z)^{p+5}}{2\chi\delta^{p+2}}\left(\frac{D_{\rm L}}{\rm Gpc}\right)^{-1}\left(\frac{\Phi}{\rm mas\cdot GHz}\right)^{-1}\times \\ \ \\
\displaystyle\times\left(\frac{\nu_{\rm obs}}{\rm GHz}\right)^2\left(\frac{T_{\rm b,\,obs}}{10^{12}{\rm K}}\right)^{p+2}.
\end{array}\label{neq3}
\end{equation}
For $p=2$}
\begin{equation}
\begin{array}{l}
\displaystyle \left(\frac{n_{\rm rad}}{{\rm cm}^{-3}}\right)=8.2\cdot 10^{3}\frac{\Gamma\sin^2\varphi(1+z)^7}{2\chi\delta^4}f(2)\times \\ \ \\
\displaystyle\times\left(\frac{D_{\rm L}}{\rm Gpc}\right)^{-1}\left(\frac{\Phi}{\rm mas\cdot GHz}\right)^{-1}\left(\frac{\nu_{\rm obs}}{\rm GHz}\right)^2\left(\frac{T_{\rm b,\,obs}}{10^{12}{\rm K}}\right)^{4}.
\end{array}\label{neq3}
\end{equation}


\subsection{Physical parameters in the sources with extreme brightness temperatures}

The above estimates we can apply to two objects with measured brightness temperature and core-shift.
The equations (\ref{Eq0}), (\ref{Seq3}) and (\ref{neq3}) permit us to obtain estimates for the radiating particles magnetization $\Sigma$,
magnetic field $B$ and radiating particles number density $n_{\rm rad}$ in the observed radio core (radiating domain) if we have precise enough measurement of the
brightness temperature. On the other hand, if we have the lower limit for the brightness temperature \citep{Lobanov-15}, these expressions
provide the lower limit for particle number density $n_{\rm rad}$ and upper limits
for the magnetic field $B$ and magnetization parameter $\Sigma$.

We will calculate the magnetic field $B$, particle number density (in nucleus frame) $n_{\rm rad}$ and magnetization (measure of equipartition)
$\Sigma$ for the blazars BL Lac and 3C273 basing on the measurements of the core brightness temperature by \citet{Gomez-16} and \citet{Kovalev-2016}.
The other parameters we need are the Doppler factor, Lorentz factor of a flow, the observation angle $\varphi$, red shift $z$, and the
half-opening angle $\chi$. The red shift and apparent velocity
\begin{equation}
\beta_{\rm app}=\frac{\beta\sin\varphi}{1-\beta\cos\varphi} \label{beta_app}
\end{equation}
we take from \citet{Lister-13}. 

\nv{There are several approaches to deduction of a Doppler factor $\delta$ from the observed jet parameters.
The first one employs the relation $\varphi\approx\gamma^{-1}$ and provides $\delta_{\beta_{\rm var}}=\beta_{\rm app}$. 
The modeling of a probability of a source having 
a Doppler factor $\delta=\beta_{\rm app}$ from the flux density-limited sample \citep{Cohen-07}
shows that this probability is peaked around unity for a large sample. 
Another assumption used in this method is that the
pattern speed is approximately equal to the flow speed, and the results of modeling by \citet{Cohen-07} support it.
The second way to estimate the jet Doppler factor is based on the assumption that the characteristic
time of variability of a bright knot in a jet gives us information about the light-travel time across the
knot of the observed angular size. This allows to calculate the variability Doppler factor \citet{Jorstad-05}.
It has been shown by \citet{Jorstad-05} for the set of 15 sources that $\delta_{\beta_{\rm app}}$ and $\delta_{\rm var}$ correlate with each other, 
following approximately the linear dependence $\delta_{\beta_{\rm app}}\approx 0.72 \delta_{\rm var}$. This 
supports a possibility of using $\delta_{\beta_{\rm var}}$ as an estimate for the Doppler factor of each individual source.
The third method used by \citet{Hovatta-09} is based on comparison of the variability brightness temperature
to the equipartition brightness temperature.}   

\nv{For the two sources with extreme brightness temperature the Doppler factor can be estimated by the first two methods.
We do not use the results by \citet{Hovatta-09}, since these have been obtained using the equipartition assumption. 
For 3C 273 source $\delta_{\beta_{\rm app}}=14.86$ \citep{Lister-13} and $\delta_{\rm var}=12.6$ \citep{Jorstad-05}.
For BL Lac $\delta_{\beta_{\rm app}}=9.95$ \citep{Lister-13} and $\delta_{\rm var}=8.1$ \citep{Jorstad-05}. 
Here we have chosen the
maximal value for $\delta_{\rm var}$ from the set of different values for different knots. Both methods provide the 
estimates for the Doppler factors which are in good agreement with each other. 
For our purposes we use the estimate $\delta_{\beta_{\rm app}}$, as for the sources under consideration 
$\delta_{\beta_{\rm app}}>\delta_{\rm var}$, thus providing the upper limit for the values of $B$ and $\Sigma$ and the lower limit for 
$n_{\rm rad}$ --- the closest to the equipartition values limits.}

The expression for the observation angle can be found using the Doppler factor definition and equation (\ref{beta_app}):
\begin{equation}
\varphi={\rm atan}\left(\frac{2\beta_{\rm app}}{2\beta_{\rm app}^2-1}\right).
\end{equation}
We also use the observations of apparent half-opening angle by \citet{Push-09}. Knowing the observation angle $\varphi$ and apparent 
half-opening angle $\chi_{\rm app}$ one can obtain the half-opening angle
\begin{equation}
\chi\approx\chi_{\rm app}\sin\varphi/2.
\end{equation} 
The luminosity distance $D_{\rm L}$ obtained according to the $\Lambda$ dark matter cosmological model with
$H_0=71\;{\rm km\,s^{-1}\,Mpc^{-1}}$, $\Omega_{\rm m}=0.27$ and $\Omega_{\Lambda}=0.73$ \citep{Komatsu-09}. 

{\it BL Lac parameters.} For this object we use the brightness temperature measurements by \citet{Gomez-16}. We choose the measurement at $\nu_{\rm obs}=15$~GHz,
as this frequency is closest to the frequencies used to estimate a core-shift for this object. The lower estimate for the observed
brightness temperature is $7.9\cdot 10^{12}$~K. We also employ the following observable parameters needed to obtain the physical
properties at the core of BL Lac. They are: $z=0.069$, $\beta_{app}=9.95$ \citep{Lister-13}, $\chi_{\rm app}=26.2^{\circ}$ \citep{Push-09}, 
and $\Phi=0.55\;{\rm mas}\cdot{\rm GHz}$ \citep{Push-12}. From these we find $D_{\rm L}=0.31$~Gpc, $\varphi=0.1$, $\chi=0.02$, and $\Gamma\approx 20$.
Substituting these parameters into (\ref{Eq0}), (\ref{Seq3}) and (\ref{neq3}), we obtain:
$B=3.3\cdot 10^{-2}$~G, $n_{\rm rad}=3.4\cdot 10^{7}\;{\rm cm}^{-3}$, $\Sigma=1.3\cdot 10^{-5}$. 

{\it 3C273 parameters.} For this object we take the measurements of the brightness temperature by \citet{Kovalev-2016} at $\nu_{\rm obs}=4.8\;{
\rm GHz}$. For this object the observed brightness temperature is $T_{\rm b,\,obs}=13\cdot 10^{12}$~K.
We employ the following observable parameters for the 3C273: $z=0.158$, $\beta_{app}=14.86$ \citep{Lister-13}, 
$\chi_{\rm app}=10.0^{\circ}$ \citep{Push-09}, and $\Phi=0.34\;{\rm mas}\cdot{\rm GHz}$ \citep{Push-12}.
From these we calculate $D_{\rm L}=0.75$, $\varphi=0.067$, $\chi=0.006$, and $\Gamma\approx 30$.
For these parameters we obtain the following physical parameters:
$B=8.1\cdot 10^{-3}$~G, $n_{\rm rad}=1.4\cdot 10^{7}\;{\rm cm^{-3}}$, $\Sigma=2.9\cdot 10^{-6}$.


\nv{The magnitudes of magnetic field and particle number density in radiation region, obtained basing on the brightness temperature measurements,
differ significantly from the jet parameters, obtained by \citet{Lobanov-98, OSG-09, Hirotani-05} basing on the equipartition assumption.}
From the equations (\ref{Eq0}), (\ref{Seq3}) and (\ref{neq3}) we can check the would-be observed brightness temperature for the system in equipartition
and, consequently, the equipartition magnetic field $B_{\rm eq}$ and radiating particle number density $n_{\rm rad,\,eq}$ for our sources. Setting for each source
$\Sigma=1$, which corresponds to the equipartition regime, we obtain for BL Lac $T_{\rm b,\,eq}=1.9\cdot 10^{12}$~K, and for 3C 273 $T_{\rm b,\,eq}=6.8\cdot 10^{11}$~K.
The equipartition magnetic field and radiating particles number density in the observed core are for BL Lac are $B_{\rm n,eq}=0.56$~G and
$n_{\rm rad,eq}=1.2\cdot 10^{5}\;{\rm cm}^{-3}$, and for 3C 273 are $B_{\rm n,eq}=3$~G and
$n_{\rm rad,eq}=26\;{\rm cm}^{-3}$.
\nv{The extreme values of the physical parameters of the radiating \nv{region} are in accordance with the conclusions by \citet{Readhead-94}, who had found that even the brightness temperatures at Compton catastrophe limit would need an extreme departures from the equipartition. 
Here we want to mention that the 3C 273 source demonstrates also the extreme magnitude of Michel's magnetization,
even calculated basing on the equipartition assumptions \citep{NBKZ-15}.}

We have obtained the radiating particles magnetization for the two sources with extreme brightness temperatures
to be of the order of $10^{-5}$. The obtained radiating magnetization allows us to estimate the 
total outflow magnetization $\sigma_{\rm tot}$. Indeed, the total magnetization is defined as  
a function of a bulk flow magnetization $\sigma\approx 1$ for MHD outflows and radiating magnetization $\Sigma\ll 1$:
\begin{equation}
\sigma_{\rm tot}=\frac{B^2}{4\pi mc^2 n\Gamma+4\pi mc^2 n_{\rm rad}\ln\frac{\gamma_{\rm max}}{\gamma_{\rm min}}/\Gamma}=\frac{1}{1/\sigma+1/\Sigma}.
\end{equation} 
Thus, we conclude that the radiating plasma must be highly relativistic so as to dominate the particle
energy flux, at least in the radiation domain, so that the total outflow magnetization
\begin{equation}
\sigma_{\rm tot}\approx\Sigma\ll 1.
\end{equation}

The non-equipartition physical parameters in the core have extreme values. Indeed, let us estimate the
maximum particle number density in a jet provided that the total jet power is in particles kinetic energy.
Thus, 
\begin{equation}
\Gamma mc^3\int_0^{R_{\rm j}}n^{\rm lab}(r_{\perp})2\pi r_{\perp}dr_{\perp}\le P_{\rm jet}.
\end{equation} 
For the uniform transversal number density distribution we get 
\nv{
\begin{equation}
\left(\frac{n^{\rm lab}}{{\rm cm^{-3}}}\right)\le 10^4\left(\frac{P_{\rm jet}}{10^{45}\;{\rm erg/s}}\right)\left(\frac{R_{\rm jet}}{0.1\;{\rm pc}}\right)^2.\label{n_lim}
\end{equation}
Although the total jet power is not always known, the estimate based on correlation between the total jet power and
radio power \citep{Cavagnolo-10} may be applied.
This provides for both sources the values of $P_{\rm BL Lac}\approx 1.2\cdot 10^{44}$ erg/s and $P_{\rm 3C 273}\approx 3.5\cdot 10^{45}$ erg/s \citep{NBKZ-15}.
For the inequality (\ref{n_lim}) to hold for the obtained values of $n_{\rm rad}$, the jet radius $R_{\rm jet}$ has to be approximately 20~pc and 2~pc respectively.
These values exceed the measured jet radius for M87 \citep{Mertens-16} of the order of 0.1~pc.}
This may mean that we underestimate the magnetic field amplitude, or that the physical conditions in the radiating domain are very different from the
conditions over the larger jet domain, so that the Blandford--K\"onigl model is not applicable for the radiating core.
Below we will address the first issue of probable underestimation of the magnetic field magnitude.


\section{The simplest non-uniform model}
\label{sec:non-uni}

We see that the standard approach of Blandford--K\"onigl model applied for the observed extreme brightness temperatures gives
the small magnetic field and unphysically high particle number density. However, as it has been pointed out by \citet{Marscher-77}, 
non-uniform models with transversal structure provide the strong dependance of physical parameters of a flow 
on observables. In this section we will relax the assumption of a uniform distribution of a magnetic field across the radiating domain. 
Here we will employ the MHD model for the transversal jet structure in the radiating domain in order to calculate the 
spectral flux and thus obtain the expression for the magnetic field as a function of an observed brightness temperature. 
We plan to reconsider the effect of non-uniform distribution
of physical parameters on core-shift effect in the future paper. 

\subsection{Model with the uniform velocity across the jet}

We assume the radiation site being the part of a continuous cylindrical jet with the bulk Lorentz-factor $\Gamma$ with plasma excited by some process 
so it has a power-law energy distribution (\ref{dn1}) in the jet frame (which in our model is a pattern frame also). 
We assume a \nv{radiating region of a jet} being uniform along the jet axis, but having a transversal structure:
magnetic field $B(r_{\perp})$ and particle number density $n(r_{\perp})$ are functions of the radial distance from the jet axis $r_{\perp}$. These we will specify
below in the text.

\nv{Modeling the transversal jet structure needs solving the MHD equations --- the Grad-Shafranov equation together with the Bernoulli equation (see e.g. review
by \citet{Beskin-10}). In general, these cannot be solved analytically, although in some special cases (self-similarity or special geometry) the solution may be obtained.
The numerical MHD simulations provide a powerful instrument in constructing the jet internal structure models. In this work we will use
the obtained earlier analytical and numerical results as a simplest model for the relativistic jet transverse structure.}
The particle number density and toroidal magnetic field dependance on the distance from the jet axis
we will model by two domains. The first one is a jet central core, which we define as a central part of jet with 
uniform $n$ and $B_{\varphi}$ distributions \citep{Kom-07, Lyub-09, NBKZ-15, BTch-16}. The size of a central core $R_{\rm c}$ is of the order of a few 
light cylinder radii. In particular, the numerical modeling by \citet{BTch-16} gives for $R_{\rm c}\approx R_{\rm L}$, and
semi-analytical modeling by \citep{Kom-07, BN-09, NBKZ-15}  gives $R_{\rm c}\approx 5 R_{\rm L}$. As these results are very close,
we will use for simplicity $R_{\rm c}=R_{\rm L}$.   
Further, the same modeling allow us to approximate the particle number density and the poloidal and toroidal magnetic fields in a jet in
the second domain by the power laws \citep{NBKZ-15, BTch-16}. So, we will use the following functions as an approximation for $B(r_{\perp})$ and $n(r_{\perp})$:   
\begin{equation}
n_{\rm rad}=n_0\left\{
\begin{array}{rl}
1,& r_{\perp}\le R_{\rm L}, \\ \ \\
\left(R_{\rm L}/r_{\perp}\right)^2,& R_{\rm L}<r_{\perp}\le R_{\rm j},
\end{array}
\right.\label{n_scale}
\end{equation} 
\begin{equation}
B_{\rm P}=B_0\left\{
\begin{array}{rl}
1,& r_{\perp}\le R_{\rm L}, \\ \ \\
\left(R_{\rm L}/r_{\perp}\right)^2,& R_{\rm L}<r_{\perp}\le R_{\rm j},
\end{array}
\right.\label{Bp_scale}
\end{equation}
\begin{equation}
B_{\varphi}=B_0\left\{
\begin{array}{rl}
r_{\perp}/R_{\rm L},& r_{\perp}\le R_{\rm L}, \\ \ \\
R_{\rm L}/r_{\perp},& R_{\rm L}<r_{\perp}\le R_{\rm j},
\end{array}
\right.\label{Bphi_scale}
\end{equation}

For the jet radiating \nv{region} having the flat Lorentz factor $\Gamma$ distribution across a jet, the poloidal magnetic field 
does not change with transformation from the jet into observer's frame,
and the toroidal magnetic field transforms as
\begin{equation}
B'_{\varphi}=B_{\varphi}/\Gamma.
\end{equation}
Within this model the poloidal magnetic field dominates the toridal for $r_{\perp}<\Gamma R_{\rm L}$, and
we have the following scalings for the particle number density and a magnetic field in a fluid frame:
\begin{equation}
B'=B_0 f_{\rm B}(r_{\perp})=B_0\left\{
\begin{array}{rl}
1,& r_{\perp}\le R_{\rm L}, \\ \ \\
\left(R_{\rm L}/r_{\perp}\right)^2,& R_{\rm L}<r_{\perp}\le \Gamma R_{\rm L},\\ \ \\
R_{\rm L}/r_{\perp}\Gamma,& \Gamma R_{\rm L}<r_{\perp}\le R_{\rm j},
\end{array}
\right.\label{Bprime}
\end{equation}
\begin{equation}
n'_{\rm rad}=\frac{n_0}{\Gamma}f_{\rm n}(r_{\perp})=\frac{n_0}{\Gamma}\left\{
\begin{array}{rl}
1,& r_{\perp}\le R_{\rm L}, \\ \ \\
\left(R_{\rm L}/r_{\perp}\right)^2,& R_{\rm L}<r_{\perp}\le R_{\rm j}.
\end{array}
\right.\label{nprime}
\end{equation}
Here we also assumed that the ratio of radiating particles to all the particles in a jet is constant across the jet.

The photon emission rate (\ref{rho1}) and effective absorption coefficient (\ref{ae1}) has been obtained by
\citep{GS-64, BG-70} for the randomly oriented magnetic field. The direction of a magnetic field in
the derivation in \citep{BG-70} sets the possible distribution of a pitch angle $\alpha$ of
radiating particles. In particular, for randomly oriented field, the pitch angle has a flat distribution, which gives
after averaging over $\alpha$ the appropriate factor in function $a(p)$ in (\ref{rho1}). However, we can use the 
expressions (\ref{rho1}) and (\ref{ae1}) even for the ordered magnetic field, but
randomly oriented orbits of the radiating particles, provided the pitch angle $\alpha$ also has a flat distribution. 

To obtain the numerical values, we need estimates for $R_{\rm L}$ and $R_{\rm j}$.
Further on we use the following dimension parameters for a central engine and an outflow angular velocity: the 
gravitational radius for a black hole with $M_{\rm BH}=10^9\; M_\odot$ is $r_{\rm g}=10^{-4}$~pc. We also use the result
obtained by \citet{Zam-14} for the light cylinder radius, which can be rewritten as:
\begin{equation}
\frac{\Omega_{\rm F}r_{\rm g}}{c}=\frac{2\pi\eta}{50},
\end{equation}
where $W_{\rm tot}=\eta\dot{M}c^2$. Setting $\eta=1$ we get $R_{\rm L}\approx 10r_{\rm g}$, the result, that we will use.
As to jet radius $R_{\rm j}$, the observations of M87 provide the value $R_{\rm j}\approx 0.1$~pc \nv{\citep{Mertens-16}}, so we set $R_{\rm j}=10^2 R_{\rm L}$. 

\subsection{Limiting parameters}

In the Section~\ref{s:BK} we have obtained, that the upper limit for the particle number density 
in the model with the uniform particle number density distribution is \nv{approximately $10^4\;{\rm cm^{-3}}$, assuming 
the jet parameters of the order of $P_{\rm jet}\approx 10^{45}$~erg/s and $R_{\rm jet}\approx 0.1$~pc.} 
For the non-uniform radial distribution (\ref{n_scale}) the upper limit on the particle number 
density amplitude $n_0^{\rm lab}$ is
$n_0^{\rm lab}\le 10^7$, with $n^{\rm lab}(R_{\rm j})$ being of the order of $10^3\;{\rm cm^{-3}}$.

The same bounding limits can be obtained for the toroidal magnetic field --- the field that in MHD models
defines the Poynting flux transported by a jet:
\begin{equation}
\frac{c}{4\pi}\int_0^{R_{\rm j}}B^2_{\varphi}(r_{\perp})2\pi r_{\perp}dr_{\perp}\le P_{\rm jet}.
\end{equation} 
For the uniform model $B_{\varphi}\le 1$~G. For the toroidal magnetic field defined by (\ref{Bphi_scale}) we
obtain for the field amplitude $B_0\le 40$~G for the same jet parameters. 
  
\subsection{Optical depth for small viewing angles}

Let us determine the optical depth 
\begin{equation}
\tau=\int_0^{s'_0}\ae'_{\nu'}ds'
\end{equation} 
of the radiating domain depending on $n_0$, $B_0$ and $\nu_{\rm obs}$ for the jets
directed almost at the observer --- the result applicable for the \nv{BL Lac and quasar type} sources. Since the optical depth 
is a Lorentz invariant, we will calculate it in the
fluid frame. However, we express it as a function of amplitudes of particle number density $n_0$, 
magnetic field $B_0$ in the nucleus frame, and frequency $\nu_{\rm obs}$ in the observer frame, using
the transformations from the jet frame into nucleus or observer frame (\ref{nu_nupr}) and (\ref{Bprime})--(\ref{nprime}).  
For small viewing angles $\varphi\ll 1$ we simplify the integration by taking $ds'\approx dz'$, so that
\begin{equation}
\begin{array}{l}
\displaystyle\tau(z,r_{\perp})=0.28\frac{1}{f(p)}\left(\frac{\delta}{1+z}\right)^3\left(\frac{n_0}{\rm cm^{-3}}\right)
\left(\frac{B_0}{\rm G}\right)^2\times \\ \ \\
\displaystyle\times\left(\frac{\nu_{\rm obs}}{\rm GHz}\right)^{-3}\left(\frac{z}{R_{\rm L}}\right)f_{\rm n}(r_{\perp})f_{\rm B}^2(r_{\perp}).
\end{array}\label{tau1}
\end{equation} 
The expression for an optical depth $\tau$ (\ref{tau1}) can be rewritten through dimensionless $\tau_0$ which depends only on the 
intrinsic radiating domain parameters and the Doppler factor
\begin{equation}
\tau_0=0.28\frac{1}{f(p)}\left(\frac{\delta}{1+z}\right)^3\left(\frac{n_0}{\rm cm^{-3}}\right)
\left(\frac{B_0}{\rm G}\right)^2\left(\frac{\nu_{\rm obs}}{\rm GHz}\right)^{-3},\label{tau0}
\end{equation}
and the dimensionless `position' factor, so
\begin{equation}
\tau=\tau_0\frac{z}{R_{\rm L}}f_{\rm n}(r_{\perp})f^2_{\rm B}(r_{\perp}).\label{tau2}
\end{equation}

Since the jet physical parameters change significantly across the jet cross-section, the optical depth of the different
domains may be greater or smaller than the unity. For example, let us describe the position 
of a surface with an optical depth equal to unity along the jet as a function of radial distance from the jet axis $r_{\perp}$.
Let us take the observed frequency $\nu_{\rm obs}=10$~GHz characteristic for the radio interferometric observations, and $\delta\approx 10$,
$\Gamma\approx 10$. 
For all reasonable parameters of a jet $n_0$ and $B_0$, the surface of $\tau=1$ is situated at geometrical depth $z$ being only
a small fraction of a parsec in the central part of a jet. For greater $r_{\perp}$ the geometrical depth of the surface $\tau=1$ 
grows extremely fast towards the jet edges. The result
depends strongly on physical parameters in a jet. For the limiting parameters (the maximal optical thickness),
if we assume $n_0=10^7\quad{\rm cm^{-3}}$ and $B_0=40$~G,
the surface $\tau=1$ for the whole jet cross-section remains optically thick 
for the radiating domain depth $z$ greater than $10^{-3}$~pc. 

However, for the less extreme parameters the situation is quite different.
If we take $n_0\approx 10^3\quad{\rm cm^{-3}}$ and $B_0\approx 1$~G --- the equipartition parameters for the uniform model obtained by \citet{Lobanov-98}, ---
the position of the surface with $\tau=1$ must be of the order of a few parsec at $r_{\perp}=\Gamma R_{\rm L}$, which means, that
the radiating domain is optically thick in the central jet part and optically thin at the outer jet domain.

\subsection{Non-uniform jet velocity}

\nv{There are some observational indications of a non-flat transversal Lorentz factor structure:
the limb-brightening (see e.g. \citet{Giroletti-08}) and M87 observed velocity transverse profile \citep{Mertens-16}.
In the latter work there have been detected super-luminal velocities in the limbs as well as in the central stream.
The numerical \citep{Tch-08, Tch-09} and analytical \citep{BN-06, Lyub-09} modeling 
show that the bulk flow Lorentz factor is not constant in the transversal jet direction.
In particular, the following transverse Lorentz factor structure has been predicted by the 
MHD modeling:} 
\begin{equation}
\Gamma(r_{\perp})=\gamma(r_{\perp})\sigma_{\rm M}=\left\{
\begin{array}{rl}
\gamma_{\rm in}\approx 1,& r_{\perp}\le R_{\rm L}, \\ \ \\
r_{\perp}/R_{\rm L},& R_{\rm L}<r_{\perp}\le \sigma_{\rm M} R_{\rm L},\\ \ \\
\sigma_{\rm M},& r_{\perp}>\sigma_{\rm M} R_{\rm L}.
\end{array}
\right.
\end{equation}
Here $\sigma_{\rm M}$ is Michel's magnetization parameter --- the ratio of Poynting flux to particles kinetic energy flux
at the base of an outflow. It bounds the maximum Lorentz-factor as $\Gamma<\sigma_{\rm M}$.
We are using the dependances (\ref{n_scale})--(\ref{Bphi_scale}) for a particle number density and a magnetic field.

The drift bulk velocity of plasma has both a toroidal $v_{\rm dr,\varphi}=v_{\rm dr}B_{\rm P}/B_{\varphi}$ and a poloidal 
$v_{\rm dr,\,P}=v_{\rm dr}B_{\varphi}/B_{\rm P}$ components. However, as outside the light cylinder $R_{\rm L}=\Omega_{\rm F}/c$ the 
toroidal magnetic field is much greater than the poloidal, we will neglect by the latter. Thus, the poloidal magnetic field 
does not change with transformation from the jet into observer's frame for $r_{\perp}>R_{\rm L}$,
and the toroidal magnetic field transforms as
\begin{equation}
B'_{\varphi}=B_{\varphi}/\Gamma(r_{\perp}).
\end{equation}
Inside the light cylinder we have the opposite: we transform the poloidal magnetic field, and the toroidal field is unchanged. 
So, under these assumptions in the fluid frame the toroidal magnetic field dominates the poloidal one at 
$r_{\perp}>R_{\rm L}$, and we have the following magnetic field and particle number density transversal profiles in the
jet frame:
\begin{equation}
B'=B_0\left\{
\begin{array}{rl}
1,& r_{\perp}\le R_{\rm L}, \\ \ \\
\left(R_{\rm L}/r_{\perp}\right)^2,& R_{\rm L}<r_{\perp}\le \sigma_{\rm M}R_{\rm L},\\ \ \\
R_{\rm L}/r_{\perp}\sigma_{\rm M},& \sigma_{\rm M}R_{\rm L}<r_{\perp}\le R_{\rm j},
\end{array}
\right.\label{B_prime_scale}
\end{equation}
\begin{equation}
n'_{\rm rad}=n_0\left\{
\begin{array}{rl}
1,& r_{\perp}\le R_{\rm L}, \\ \ \\
\left(R_{\rm L}/r_{\perp}\right)^3,& R_{\rm L}<r_{\perp}\le \sigma_{\rm M}R_{\rm L},\\ \ \\
\left(R_{\rm L}/r_{\perp}\right)^2/\sigma_{\rm M},& \sigma_{\rm M}R_{\rm L}<r_{\perp}\le R_{\rm j}.
\end{array}
\right.\label{n_prime_scale}
\end{equation}

The flow Doppler factor depends on the distance from the axis as well. We introduce
the Doppler factor for the fastest part of a flow $\delta_0=1/\sigma_{\rm M}(1-\beta(r_{\perp})\cos\theta)$. As the flow is relativistic,
we neglect by the change in $\beta$ across the flow, and use
\begin{equation}
\delta(r_{\perp})=\frac{\delta_0}{\gamma(r_{\perp})}.
\end{equation}
Due to this, the observed spectral flux will be much less homogeneous than it is suggested by mere change in $B'$ and $n'$
in comparison with the model with the uniform jet velocity.

The optical thickness as a function of $z$ and $r_{\perp}$ is now given by
\begin{equation}
\begin{array}{l}
\displaystyle\tau(z,r_{\perp})=\frac{0.28}{f(p)}\left(\frac{\delta_0}{1+z}\right)^3\left(\frac{n_0}{\rm cm^{-3}}\right)
\left(\frac{B_0}{\rm G}\right)^2\times \\ \ \\
\displaystyle\times\left(\frac{\nu_{\rm obs}}{\rm GHz}\right)^{-3}
\left(\frac{z}{R_{\rm L}}\right)\frac{f_{\rm n}(r_{\perp})f_{\rm B}^2(r_{\perp})}{\gamma^3(r_{\perp})}=\\ \ \\
\displaystyle=\tau_{0,2}\frac{z}{R_{\rm L}}\frac{f_{\rm n}(r_{\perp})f_{\rm B}^2(r_{\perp})}{\gamma^3(r_{\perp})}.
\end{array}\label{tau1_gamma}
\end{equation}
This equation coincides with (\ref{tau1}) except for additional factor $\gamma^3(r_{\perp})$. This means that the optical
depth in the central jet part is the same as in a limit of a uniform Lorentz factor. However, the position of a surface $\tau=1$
grows much more rapidly for $r_{\perp}>R_{\rm L}$, and for reasonable depth $L$ of radiating domain it becomes optically thin.
The equation for surface $\tau=1$ is given by
\begin{equation}
\frac{z(r_{\perp})}{R_{\rm L}}=\frac{1}{\tau_{0,2}}
\left\{
\begin{array}{rl}
\displaystyle \sigma_{\rm M}^{-3},& r_{\perp}\le R_{\rm L},\\ \ \\
\displaystyle (r_{\perp}/R_{\rm L})^{10}\sigma_{\rm M}^{-3},& R_{\rm L}<r_{\perp}\le \sigma_{\rm M}R_{\rm L},\\ \ \\
\displaystyle (r_{\perp}/R_{\rm L})^{4}\sigma_{\rm M}^{3},& \sigma_{\rm M}R_{\rm L}<r_{\perp}\le R_{\rm j}.
\end{array}
\right.
\end{equation}
Although the radiating domain optical thickness declines more rapidly than in the case of a flat velocity distribution,
for upper limits for $n_0$ and $B_0$ the outer part of an outflow stays optically thick for $L<2\cdot 10^{-2}$~pc. 

\subsection{Observed flux}

Now we will calculate the observed spectral flux of a model radiating domain with the non-uniform distribution
of a particle number density and a magnetic field for small viewing angles. We first determine the spectral flux
in the fluid frame, where an emissivity and effective absorption are readily calculated \citep{GS-64, BG-70}, and than
transform it into the observer frame \citep{R-L-79}.

A spectral flux in the jet frame is defined by
\begin{equation}
S'_{\nu'}=\frac{1}{d^2}\int_{\Omega'}j'_{\nu'}(\nu')dV'e^{-\int \ae'_{\nu'}(\nu')ds'},
\end{equation}
where $\Omega'$ is a radiating domain. Using (\ref{rho1}) and (\ref{ae1}) for the synchrotron radiation, and
having $j'_{\nu'}(\nu')=\hbar\nu'\rho'_{\nu'}(\nu')$, we obtain for the flux in a jet frame written through the 
observed frequency $\nu_{\rm obs}$ and a particle number density $n$ and a magnetic field $B$ in the nucleus frame the following expression:
\begin{equation}
S'_{\nu}(\nu,\,n_0,\,B_0)=0.16\frac{\hbar\nu}{d^2}\frac{\nu}{r_0c}\left(\frac{\nu_{B_0}}{\nu}\right)^{-1/2}
\left(\frac{1+z}{\delta}\right)^{5/2}I,\label{flux_non_un}
\end{equation}
where the integral $I$ has dimension $\rm cm^2$ and is defined by
\begin{equation}
I=\int_0^{R_{\rm j}}\frac{1}{\sqrt{f_{\rm B}(r_{\perp})}}r_{\perp}dr_{\perp}\left[1-e^{-\tau_0\frac{L}{R_{\rm L}}f_{\rm n}(r_{\perp})f^2_{\rm B}(r_{\perp})}\right].
\end{equation}
If the whole jet cross-section is optically thick, the integral can be easily calculated.
In the inner domain $r_{\perp}\in[0,\,\Gamma R_{\rm L}]$ it is equal
\begin{equation}
I_{\rm in}=R^2_{\rm L}\left(\frac{1}{6}+\frac{\Gamma^3}{3}\right),
\end{equation} 
and in the outer domain $r_{\perp}\in(\Gamma R_{\rm L},\,R_{\rm j}]$
\begin{equation}
I_{\rm out}=\frac{2}{5}R^2_{\rm L}\left(\sqrt{\Gamma}\left(\frac{R_{\rm j}}{R_{\rm L}}\right)^{5/2}-\Gamma^3\right),
\end{equation} 
and
\begin{equation}
I\approx\frac{2}{5}\sqrt{\Gamma}\sqrt{\frac{R_{\rm j}}{R_{\rm L}}}R_{\rm j}^2,\label{form_fact}
\end{equation}
the outer radiating domain providing the major part of the total flux. 

In order to link the spectral flux in the jet frame with the observed brightness temperature,
we use the Lorentz invariance of $S_{\nu}/\nu^3$ \citep{R-L-79}. Substituting (\ref{flux_non_un}) and (\ref{form_fact})
into (\ref{S1}), one obtains:
\begin{equation}
\left(\frac{B_0}{\rm G}\right)=6.4\cdot 10^{-4}\Gamma\frac{R_{\rm j}}{R_{\rm L}}\frac{\delta}{1+z}\left(\frac{\nu_{\rm obs}}{\rm GHz}\right)
\left(\frac{T_{\rm b,obs}}{10^{12}\,\rm K}\right)^{-2}.\label{Bres}
\end{equation} 
Compare this result with the uniform model (\ref{Eq0}). The non-uniform model of optically thick outflow gives for a magnetic field the amplitude
of the uniform model multiplied by a ``geometrical'' factor $R_{\rm j}/R_{\rm L}$, which raise the value
by two orders. We see that the uniform model underestimates even the average value of a magnetic field in comparison with the non-uniform model. Indeed, 
$B_0$ is greater than the uniform magnetic field everywhere across the jet, and for
$R_{\rm j}/R_{\rm L}=10^2$ both fields become comparable only at the jet boundary.

The magnetic field estimated in the frame of a model with non-uniform jet velocity distribution obeys the same expression
(\ref{Bres}), since the outer domain $r_{\perp}>\sigma_{\rm M}R_{\rm L}$  contributes most in a spectral flux, and the velocity profile
in this domain is the same for two models.

For the two sources with the measured extreme brightness temperatures, the magnetic field estimated by the non-uniform model is:
\begin{equation}
B_{\rm non-uni}^{\rm BLLac}=3\;{\rm G},\quad B_{\rm non-uni}^{\rm 3C273}=0.7\;{\rm G}.\label{Bnonuni}
\end{equation}
   
\section{Astrophysical applications and discussions}

In the frame of Blandford--K\"onigl model we have rederived the expressions for a magnetic field and a particle number density
used by \citep{Zdz-15} as a tool to estimate these physical parameters of radiation domain in a jet independently
of the equipartition assumption. However, contrary to their work, we expressed the parameters through the brightness 
temperature. As the values for $T_{\rm b,\,obs}$ obtained with high resolution exceed by two orders the equipartition 
temperature derived by \citet{Readhead-94}, the jet parameters differs from the ones corresponding to equipartition, the measure of
equipartition being of the order of $10^{-6}\div 10^{-5}$.
In particular, the magnetic field in radiating domain has an order of $10^{-3}$, which according to (\ref{BK_model}) provides the magnetic
field at the gravitational radius
$B_{\rm g}$ of the order of a few G. The expression for $n$ gives the unphysically high amount of particles of the order of $10^7\;{\rm cm^{-3}}$, 
since such an amount would carry energy exceeding the total jet power. However, the two sources regarded in this work
have core-shifts smaller than typical errors estimated in \citet{Push-12} of $0.05$~mas. Thus, the results for $\Sigma$ 
and $n$ may be subject to big errors. 

We have obtained the expression for the magnetic field amplitude $B_0$ that can be estimated by measurement of a 
brightness temperature. The expression is applicable for blazars, since it uses the head-on model of
radiation transfer for the non-uniform cylindrical optically thick radiation domain with profiles for
particle number density and magnetic field distribution based on MHD modeling. 
The field amplitude characterizes the radiating core region only and may differ from the
other domains along the jet. The expression for $B_0$ differs form the expression for the homogeneous
model of the radiating domain by the factor $R_{\rm j}/R_{\rm L}$, which gives two orders of magnitude.

In the frame of MHD models the amplitude $B_0$ characterizes both poloidal and toroidal magnetic field. Thus,
this amplitude provides us with an instrument of checking the electrodynamic model of the black hole energy
extraction. Indeed, if we assume the unipolar inductor model for the AGN,
the total jet power is given by \citep{Beskin-10}
\begin{equation}
P_{\rm tot}=\left(\frac{\Omega r_{\rm g}}{c}\right)^2 B_{\rm g}^2 r_{\rm g}^2 c.
\end{equation}
As we can estimate $B_0$ as being of the order of $1$~G, using (\ref{Bp_scale}), we can correlate the magnetic field amplitude with the total flux
crossing the gravitational radius, and thus obtain the poloidal magnetic field at the base of an outflow $B_{\rm g}$.
Indeed, on the one hand,
\begin{equation}
\Psi_{\rm tot}=\pi B_{\rm g}r_{\rm g}^2,
\end{equation}
and on the other hand
\begin{equation}
\Psi_{\rm tot}\approx 2\pi B_0 R_{\rm L}^2\ln\frac{R_{\rm j}}{R_{\rm L}}.
\end{equation}
From these equalities we have
\begin{equation}
B_{\rm g}=2B_0\ln\frac{R_{\rm j}}{R_{\rm L}}\left(\frac{R_{\rm L}}{r_{\rm g}}\right)^2,\label{Brg}
\end{equation}
which gives $B_{\rm g}\approx 10^3$. It is about an order smaller than the Eddington magnetic field (see, e.g., \citet{Beskin-10})
\begin{equation}
B_{\rm Edd}=10^4\left(\frac{M_{\rm BH}}{10^9 M_{\odot}}\right)^{-1/2}~{\rm G}.
\end{equation} 
Such a magnitude for $B_{\rm g}$ provides within the electrodynamical model for the total jet power an estimate $P_{\rm tot}\approx 3\cdot 10^{43}$~erg/s.

The above expression (\ref{Brg}) for a poloidal magnetic field magnitude is consistent with the 
maximum possible amount of magnetic flux achieved by magnetically arrested disks. Indeed, for 
$M_{\rm BH}=10^9 M_{\odot}$ and accretion rate 10\% of Eddington luminosity \citep{Hawley-15}, 
$\Phi_{\rm MAD}=3\cdot 10^{33}\;{\rm G\,cm^2}$, and for $B_{\rm g}$ given by (\ref{Bnonuni}) and (\ref{Brg})
the magnetic flux is of the order of $10^{33}\;{\rm G\,cm^2}$.


\section*{Acknowledgments}
\nv{The author thanks the anonymous referee for
the suggestions which helped to improve the paper.}
This research has made use of data from the MOJAVE
data base that is maintained by the MOJAVE team \citep{Lister-09}.
This work was supported by the Russian Foundation for Basic Research, 
grant 16-32-60074\_mol\_a\_dk.

\end{document}